\documentclass[11pt]{article} 

\usepackage{xcolor}

\usepackage[normalem]{ulem}
\newcommand{\stkout}[1]
{\ifmmode\text{\sout{\ensuremath{#1}}}\else\sout{#1}\fi}
\def\ci{\perp\!\!\!\perp}

\usepackage[a4paper, margin=1in]{geometry} % Standard margins for A4

\usepackage{parskip}
\usepackage[T1]{fontenc}       % Proper font encoding for better output
\usepackage[utf8]{inputenc}    % UTF-8 encoding for special characters
\usepackage{lmodern}           % Improved font rendering (Latin Modern)
\usepackage{microtype}         % Enhances typography for better spacing

\usepackage{amsmath, amssymb, amsthm}  % Core math packages
\usepackage{mathtools}                 % Enhancements for amsmath
\usepackage{bbm}                        % Indicator function and blackboard bold symbols
\usepackage{bm}                         % Bold math symbols
\usepackage{commath}                    % Provides \dif for derivatives and other useful macros
\usepackage{nicefrac}                   % Compact fractions
\usepackage{xfrac}                      % Alternative fraction styles

\usepackage{mathrsfs}        % Ralph Smith’s Formal Script, provides \mathscr
\DeclareMathAlphabet{\mathscr}{OMS}{rsfs}{m}{n}  % Ensure \mathscr uses rsfs font family

\theoremstyle{plain}

\theoremstyle{definition}

\theoremstyle{remark}

\usepackage{graphicx}         % Include images
\usepackage{caption}          % Better caption handling
\usepackage{subcaption}       % Subfigures (grouped figures)
\usepackage{booktabs}         % Professional-looking tables
\usepackage{multirow}         % Merging rows in tables
\usepackage{array}            % Improved table layout
\usepackage{float}            % Better figure placement
\usepackage{wrapfig}          % Text wrapping around figures

\usepackage[authoryear]{natbib}  % Citation style (numeric)
\usepackage{hyperref}         % Clickable links in PDF
\hypersetup{
    colorlinks=true,          % Enable colored links
    linkcolor=blue,           % Color for internal links
    citecolor=blue,           % Color for citations
    urlcolor=blue             % Color for URLs
}

\usepackage{tikz}             % Core TikZ package for graphics
\usetikzlibrary{
    decorations.pathmorphing, 
    positioning,              % Easier node placement
    arrows,                   % Arrows for diagrams
    decorations.pathreplacing,% Curly braces for braces
    calc                     % Mathematical calculations
}

\usepackage{pgfplots}
\usepgfplotslibrary{fillbetween} % For filling areas under curves
\pgfplotsset{compat=1.17} % Compatibility setting

\usepackage{algorithm}        % Floating environment for algorithms
\usepackage{algpseudocode}    % Pseudocode syntax

\usepackage{colortbl}      % Color tables: cell shading and colored rows
\usepackage{hhline}        % Enhanced horizontal lines in tables
\usepackage{calc}          % Perform arithmetic calculations in LaTeX lengths
\usepackage{tabularx}      % Tables with adjustable-width columns
\usepackage{comment}       % Environment for commenting out blocks

\usepackage{xcolor}           % Colors for text and highlighting
\usepackage{enumitem}         % Custom bullet points and numbering
\usepackage{xspace}           % Smart spacing in custom macros
\usepackage{fancyvrb}         % Improved verbatim environment
\usepackage{tcolorbox}        % Colored boxes for notes and highlights

\usepackage{changepage}       % Add indent paragraph

\usepackage{todonotes}        % Inline and margin to-do notes, with \listoftodos

\newcommand{\red}{\textcolor{red}}

\allowdisplaybreaks        % Allow page breaks in multi-line display math

\title{Response to Discussions of ``Causal and Counterfactual Views of Missing Data Models''} 

\author{
    Razieh Nabi,$^1$ Rohit Bhattacharya,$^2$ Ilya Shpitser,$^3$ James M. Robins$^4$ \\[0.5em]
    {$^1$Department of Biostatistics and Bioinformatics, Emory University, Atlanta, GA, U.S.A.} \\ [0.15em]
    {$^2$Department of Computer Science, Williams College, Williamstown, MA, U.S.A.} \\ [0.15em]
    {$^3$Department of Computer Science, Johns Hopkins University, Baltimore, MD, U.S.A.} \\ [0.15em]
    {$^4$Departments of  Biostatistics and Epidemiology, Harvard University, Boston, MA, U.S.A.} \\ [0.35em]
}

\date{}

\begin{document}
\maketitle 

% \begin{abstract}
%     \ldots 
% \end{abstract}

%%%%%%%%%%%%%%%%%%%%%%%%%%%%%%%%%%%%%%%%%%%%%%%%%%%%%%%%%%%%
\section{Introduction}
\label{sec:intro}
%%%%%%%%%%%%%%%%%%%%%%%%%%%%%%%%%%%%%%%%%%%%%%%%%%%%%%%%%%%%

We are grateful to the discussants -- \citet{LevisKennedy2025}, \citet{LuoGeng2025}, \citet{WangVdL2025}, and \citet{YangKim2025} -- for their thoughtful comments on our paper \citep{Sinica2025}. Below we summarize our main contributions before responding to each discussion in turn. 

Graphical models have emerged as an important tool for clarifying identifying assumptions made in both causal inference \citep{pearl95seq, pearl00causality} and missing data \citep{robins97nonresponse,bhattacharya19mid, nabi20completeness, malinsky2021semiparametric, mohan2021graphical}. Our paper \citep{Sinica2025} shows how recent techniques motivated by causal graphical modeling may be fruitfully applied to obtain identification in missing data models. These recent techniques allow us to obtain novel identification results that would not be possible to obtain in standard causal inference problems,
except by imposing implausible additional assumptions, such as rank preservation.

Specifically, we formalize how identifiability of the target (i.e., complete) data law can be viewed as identification of a joint distribution over counterfactuals $L^{(1)}$,  the variables that would have been observed if all missingness indicators $R$ were set to one (that is, if no data were missing). 

This reframing yields a counterfactual analogue of the g-formula.  When assumptions encoded in the Markov properties of a missing data DAG (m-DAG) allow us to express the counterfactual g-formula in terms of the factuals, we obtain (nonparametric) identification. That is, noting that the target law $p(l^{(1)})$ satisfies $p(l^{(1)})=p(l,R=1)/p(R=1\mid l^{(1)})$, it follows that when the missingness selection model $p(R = 1\mid L^{(1)})$ is identified from the observed law, so is $p(l^{(1)})$. 

% One possible source of confusion 
In our paper we used the word ``nonparametric'' in two different ways.  The joint distribution of $(L,L^{(1)},R)$ is Markov to a given m-DAG if it factorizes as the product of  the conditional densities of each variable given its parents. In the graphical causal modeling literature, the model is said to be ``nonparametric'' just when these conditional densities are left unrestricted (with the exception that $L$ is a deterministic function of its parents).
The term ``nonparametric identification'' refers to identification in such a model.  In contrast, in the missing data literature in statistics, a model is said to be nonparametric if it places no restrictions on either the observed data law or the target data law $p(l^{(1)})$.  It is said to be nonparametric just identified (NPI) if the target law is identified from the observed data law. The permutation model \citep{robins97non-a} we discuss in our paper is known to be NPI. In contrast, all the other identified missing data models in our paper place testable restrictions on the joint distribution of the observed data. In fact, we conjecture that the permutation model is the only m-DAG model that is NPI.  In the following, we use nonparametric in the graphical causal modeling sense.

% To summarize, 
A key message of our paper is that features specific to missing data models (in particular the partial observability of the counterfactuals through the proxies $L$ and the structural restriction that $R$ and $L$ do not cause $L^{(1)}$) can deliver parameter identification in settings where analogous parameters in hidden variable causal DAGs are not identified. We catalog several such nonparametric identification techniques in m-DAGs, and we clarify similarities and differences with standard causal identification, including when additional assumptions like rank preservation might be needed for causal analogues. 

Each discussion extends or challenges our framework in important ways. \citet{LevisKennedy2025} highlight identification strategies complementary to ours, based on the existence of instrumental and shadow variables, discuss semiparametric estimation theory for functionals arising from such strategies, as well as discuss connections between m-DAGs and Single World Intervention Graphs (SWIGs). \citet{LuoGeng2025} analyze self-censoring MNAR mechanisms with binary variables, deriving identifiability results that leverage auxiliary variables. \citet{WangVdL2025} emphasize, as we do, viewing missingness as interventions, and connect our perspective to censoring in survival models. \citet{YangKim2025} and \citet{WangVdL2025} both examine challenges of applying m-DAGs in applications, particularly assumption validation, scalability, and sensitivity analysis.

%%%%%%%%%%%%%%%%%%%%%%%%%%%%%%%%%%%%%%%%%%%%%%%%%%%%%%%%%%%%
\section{Response to the discussions}

\subsection{On the discussion by Levis and Kennedy}

\cite{LevisKennedy2025} draw attention to additional causal identification tools and highlight implications for estimation. In particular, they emphasize the relevance of instrumental variables, shadow variables, and SWIGs as complementary devices for reasoning about identification in missing data problems. They also consider how identification results derived from m-DAGs can be translated into practical estimation procedures with desirable asymptotic properties. Their discussion situates our contribution within a broader pipeline of causal inference methods and points toward promising directions for future methodological development. 

We appreciate Levis and Kennedy's thoughtful remarks. On their first point, we agree that instrumental variables \citep{sun2018semiparametric}, shadow variables \citep{miao2024identification}, and related tools \citep{li2023self} are useful complements. Our emphasis in the paper, however, was on a nonparametric identification framework: aside from the Markov restrictions encoded via independence assumptions in m-DAGs, we make no further distributional assumptions. In contrast, the use of IVs or shadow variables begins in settings where nonparametric identification fails, and proceeds by imposing additional restrictions (such as functional form constraints e.g., homogeneity assumptions when using of IVs),  relevance assumptions which result in generic identification, or restrictions on the support of variables to restore identification. These extra ingredients move the analysis outside the purely nonparametric domain that was our focus.  

On estimation, we appreciate the worked example they provide. While our goal in this paper was primarily conceptual, highlighting the philosophical parallels and differences between missing data and causal identification, their discussion underscores the importance of connecting identification results to practical estimation and efficiency theory. We agree this is important future work.  

We also agree with Levis and Kennedy that causal inference problems in the presence of confounding often occur together with censoring, and that obtaining identification when both complications are present is challenging.  We also appreciate their worked example illustrating these complications.

We would like to offer two notes of caution regarding generalizing the construction presented by Levis and Kennedy, first on the appropriate generalization of the SWIG splitting construction from causal diagrams to m-DAGs, and second on the general utility of using SWIGs for obtaining identification in missing data settings.

\subsubsection*{On the translation of m-DAGs to SWIGs} 
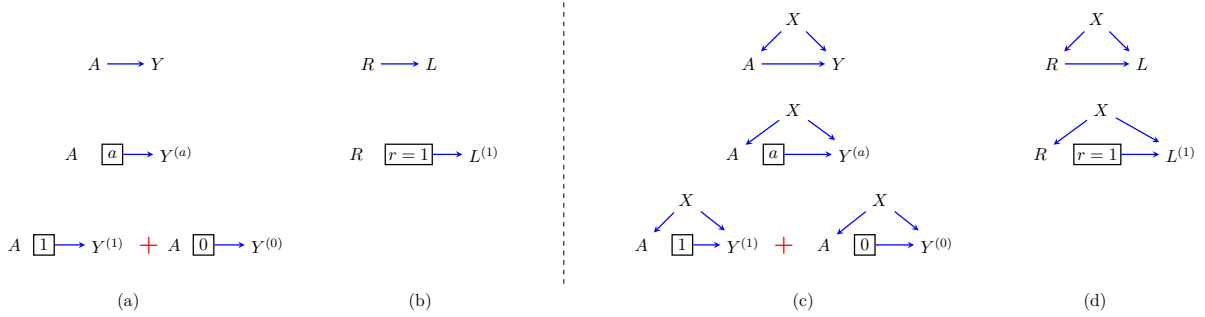
\begin{figure}[t] 
    \begin{center}
    \scalebox{0.6}{
    \begin{tikzpicture}[>=stealth, node distance=1.4cm]
        \tikzstyle{format} = [thick, minimum size=5.0mm, inner sep=0pt]
        \tikzstyle{square} = [draw, thick, minimum size=4.5mm, inner sep=2pt]
        
    % Ignorable 
    \begin{scope}[xshift=0cm, yshift=0cm]
            \path[->, thick]
            
            node[] (a) {$A$}
            node[right of=a] (y) {$Y$}

            (a) edge[blue] (y)
            ; 
        \end{scope}
        \begin{scope}[xshift=-0.5cm, yshift=-2.cm]
            \path[->, thick]
            
            node[] (a) {$A$}
            node[square, right of=a, xshift=-0.5cm] (a1) {$a$}
            node[right of=a1] (y) {$Y^{(a)}$}    

            (a1) edge[blue] (y)
            ;
        \end{scope}
        \begin{scope}[xshift=-1.75cm, yshift=-4.cm]
            \path[->, thick]
            
            node[] (a) {$A$}
            node[square, right of=a, xshift=-0.75cm] (a1) {$1$}
            node[right of=a1] (y) {$Y^{(1)}$}

            node[right of=y, xshift=-0.5cm] () {\Large{\bf \red{+}}}
            
            (a1) edge[blue] (y)
            ; 
        \end{scope}
        \begin{scope}[xshift=1.75cm, yshift=-4.cm]
            \path[->, thick]
            
            node[] (a) {$A$}
            node[square, right of=a, xshift=-0.75cm] (a1) {$0$}
            node[right of=a1] (y) {$Y^{(0)}$}
            
            (a1) edge[blue] (y)
            ; 
        \end{scope}

        % MCAR 
        \begin{scope}[xshift=6cm, yshift=0cm]
            \path[->, thick]
            
            node[] (a) {$R$}
            node[right of=a] (y) {$L$}

            node[right of=y, xshift=1.5cm, yshift=1.5cm] (d0) {}
            node[right of=y, xshift=1.5cm, yshift=-5.cm] (d1) {}
            (d0) edge[black, dashed, -, line width=0.2mm] (d1)
            
            (a) edge[blue] (y)
            ; 
        \end{scope}
        \begin{scope}[xshift=5.75cm, yshift=-2.cm]
            \path[->, thick]
            
            node[] (a) {$R$}
            node[square, right of=a, xshift=-0.25cm] (a1) {$r=1$}
            node[right of=a1, xshift=0.25cm] (y) {$L^{(1)}$}    

            (a1) edge[blue] (y) 
            ;
        \end{scope}

        % Conditionally ignorable
         \begin{scope}[xshift=14.35cm, yshift=0cm]
            \path[->, thick]
            
            node[] (a) {$A$}
            node[above right of=a, xshift=0cm] (x) {$X$}
            node[below right of=x] (y) {$Y$}

            % node[right of=y, xshift=1.75cm, yshift=1.5cm] (d0) {}
            % node[right of=y, xshift=1.75cm, yshift=-5.cm] (d1) {}
            % (d0) edge[black, dashed, -, line width=0.2mm] (d1)
            
            (a) edge[blue] (y)
            (x) edge[blue] (a)
            (x) edge[blue] (y)
            ; 
        \end{scope}
        \begin{scope}[xshift=14.cm, yshift=-2.cm]
            \path[->, thick]
            
            node[] (a) {$A$}
            node[square, right of=a, xshift=-0.5cm] (a1) {$a$}
            node[above right of=a, xshift=0.35cm] (x) {$X$}
            node[below right of=x, xshift=0.35cm] (y) {$Y^{(a)}$}
            
            (a1) edge[blue] (y)
            (x) edge[blue] (a)
            (x) edge[blue] (y)
            ; 
        \end{scope}
        \begin{scope}[xshift=12cm, yshift=-4.cm]
            \path[->, thick]
            
            node[] (a) {$A$}
            node[square, right of=a, xshift=-0.5cm] (a1) {$1$}
            node[above right of=a, xshift=0cm] (x) {$X$}
            node[below right of=x, xshift=0.25cm] (y) {$Y^{(1)}$}

            node[right of=y, xshift=-0.5cm] () {\Large{\bf \red{+}}}
            
            (a1) edge[blue] (y)
            (x) edge[blue] (a)
            (x) edge[blue] (y)
            ; 
        \end{scope}
        \begin{scope}[xshift=16cm, yshift=-4.cm]
            \path[->, thick]
            
            node[] (a) {$A$}
            node[square, right of=a, xshift=-0.5cm] (a1) {$0$}
            node[above right of=a, xshift=0.25cm] (x) {$X$}
            node[below right of=x, xshift=0.25cm] (y) {$Y^{(0)}$}
            
            (a1) edge[blue] (y)
            (x) edge[blue] (a)
            (x) edge[blue] (y)
            ; 
        \end{scope}
        
        % MAR 
        \begin{scope}[xshift=21cm, yshift=0cm]
            \path[->, thick]
            
            node[] (a) {$R$}
            node[above right of=a, xshift=0cm] (x) {$X$}
            node[below right of=x] (y) {$L$}

            % node[right of=y, xshift=2cm, yshift=1.5cm] (d0) {}
            % node[right of=y, xshift=2cm, yshift=-3.cm] (d1) {}
            % (d0) edge[black, dashed, -, line width=0.2mm] (d1)
            
            (a) edge[blue] (y)
            (x) edge[blue] (a)
            (x) edge[blue] (y)
            ; 
        \end{scope}
        \begin{scope}[xshift=20.75cm, yshift=-2.cm]
            \path[->, thick]
            
            node[] (a) {$R$}
            node[square, right of=a, xshift=-0.15cm] (a1) {$r=1$}
            node[above right of=a, xshift=0.35cm] (x) {$X$}
            node[below right of=x, xshift=0.75cm] (y) {$L^{(1)}$}
            
            (a1) edge[blue] (y)
            (x) edge[blue] (a)
            (x) edge[blue] (y)
            ; 
        \end{scope}

        \begin{scope}[xshift=0cm, yshift=-5.25cm]
            \path[->, thick]
            
            node[xshift=.75cm] (t1) {(a)}
            node[right of=t1, xshift=5.cm] (t2) {(b)}
            node[right of=t2, xshift=7.cm] (t3) {(c)}
            node[right of=t3, xshift=5.cm] (t4) {(d)}
            
            ; 
        \end{scope}
        
    \end{tikzpicture}
    }
    \end{center}
    % \vspace{-0.5cm}
	\caption{(a) Ignorable treatment model with SWIGs; (b) Its MCAR analogue, with a single relevant SWIG; (c) Conditionally ignorable treatment model with SWIGs; (d) Its MAR analogue, again with only one relevant SWIG. }
	\label{fig:swigs}
\end{figure}

For simple MAR models where the m-DAG's structure resembles that of the standard conditionally ignorable model in causal inference, the translation of an m-DAG to an equivalent SWIG is fairly natural. There are, however, important distinctions even in these simple settings. In causal inference, SWIGs provide a template indexed by treatment interventions (two templates if the treatment is binary) \citep{thomas13swig}; in the missing data analogue, only the template for $R=1$ is meaningful, since the counterfactual $L^{(0)}$ is not defined. This difference is highlighted in Figures~\ref{fig:swigs}(a, b) for a causal model satisfying ignorability and an analogous missing data model that is MCAR. Similarly, Figures~\ref{fig:swigs}(c, d) highlight the same for a conditionally ignorable causal model and a MAR missing data model.

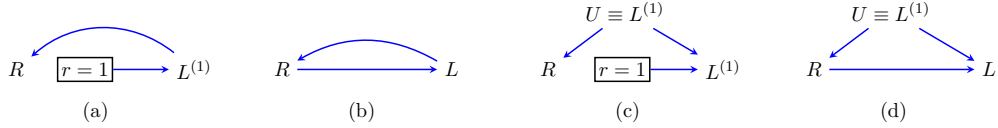
\begin{figure}[t] 
	\begin{center}
	\scalebox{0.7}{
    \begin{tikzpicture}[>=stealth, node distance=1.55cm]
        \tikzstyle{format} = [thick, minimum size=5.0mm, inner sep=1pt]
        \tikzstyle{square} = [draw, thick, minimum size=4.5mm, inner sep=2pt]
                
        \begin{scope}[xshift=0cm, yshift=0cm]
            \path[->, thick]
            
            node[] (a) {$R$}
            node[square, right of=a, xshift=-0.25cm] (a1) {$r=1$}
            node[right of=a1, xshift=0.5cm] (y) {$L^{(1)}$}
            
            (a1) edge[blue] (y)
            (y) edge[blue, bend right=40] (a)
            
            node[format, below of=a, xshift=1.5cm, yshift=0.75cm] (a) {(a)} ;
        \end{scope}

        \begin{scope}[xshift=5cm, yshift=0cm]
            \path[->, thick]
            
            node[] (a) {$R$}
            node[above right of=a, xshift=0.5cm] (x) {}
            node[below right of=x, xshift=0.5cm] (y) {$L$}
            
            (a) edge[blue] (y)
            (y) edge[blue, bend right] (a)
            
            node[format, below of=a, xshift=1.5cm, yshift=0.75cm] (b) {(b)} ;
        \end{scope}

        \begin{scope}[xshift=10.0cm, yshift=0.cm]
            \path[->, thick]
            
            node[] (a) {$R$}
            node[square, right of=a, xshift=-0.15cm] (a1) {$r=1$}
            node[above right of=a, xshift=0.35cm] (x) {$U\equiv L^{(1)}$}
            node[below right of=x, xshift=0.75cm] (y) {$L^{(1)}$}
            
            (a1) edge[blue] (y)
            (x) edge[blue] (a)
            (x) edge[blue] (y)

            node[format, below of=a, xshift=1.5cm, yshift=0.75cm] (c) {(c)} 

            ; 
        \end{scope}

        \begin{scope}[xshift=15.0cm, yshift=0.cm]
            \path[->, thick]
            
            node[] (a) {$R$}
            %node[square, right of=a, xshift=-0.15cm] (a1) {$r=1$}
            node[above right of=a, xshift=0.35cm] (x) {$U\equiv L^{(1)}$}
            node[below right of=x, xshift=0.75cm] (y) {$L$}
            
            (a) edge[blue] (y)
            (x) edge[blue] (a)
            (x) edge[blue] (y)

            node[format, below of=a, xshift=1.5cm, yshift=0.75cm] (d) {(d)} 

            ; 
        \end{scope}
        
    \end{tikzpicture}
    }
	\end{center}
    \caption{(a) A possible SWIG representation of a self-censoring missing data model; (b) Stitching SWIGs into an observed-data model produces a cycle; (c) A SWIG that draws a distinction between the full data variable $L^{(1)}$, viewed as an unobserved confounder $U$ with extra restrictions, and the observed variable $L$ under an intervention where $R$ is set to $1$; (d) The full data graph equating the full data variable $L^{(1)}$ with an unobserved confounder $U$ with special structure.
    %(c) the block-parallel MNAR model, labeling missing variables as unmeasured for the purposes of constructing a SWIG; (d) the corresponding SWIG obtained from (c).
    }
    \label{fig:mnar}
\end{figure}

Caution must be exercised when SWIGs are constructed from m-DAGs representing MNAR mechanisms.  For instance, in a missing data model with self-censoring, a natural idea for what the SWIG should look like would be the graph shown in Figure~\ref{fig:mnar}(a). However, if we try to collapse the random and fixed nodes in this SWIG to reconstruct the observed data graph,  we obtain the graph shown in Figure~\ref{fig:mnar}(b), which erroneously implies a cycle in the underlying data generating process.

The problem arises from conflating the full data variable $L^{(1)}$ in m-DAGs, and the observed data variable $L$ relabeled to be counterfactual in the SWIG construction.  An appropriate view of m-DAGs that avoids these types of difficulties is to  view full data variables such as $L^{(1)}$ as unobserved confounders $U$, with additional structure imposed by missing data consistency.  In this view, the SWIG corresponding to the self-censoring model is better represented by Figure~\ref{fig:mnar}(c), where the variable $L^{(1)}$ is viewed as an unobserved confounder $U$ with special structure, which influences both the observability indicator $R$, and the observed proxy variable $L$.  The m-DAG corresponding to this view of the self-censoring model is shown in Figure~\ref{fig:mnar}(d).

\begin{figure}[t] 
	\begin{center}
	\scalebox{0.7}{
    \begin{tikzpicture}[>=stealth, node distance=1.55cm]
        \tikzstyle{format} = [thick, minimum size=5.0mm, inner sep=1pt]
        \tikzstyle{square} = [draw, thick, minimum size=4.5mm, inner sep=2pt]

        \begin{scope}[xshift=11cm, yshift=0cm]
            \path[->, thick]

            node[yshift=1.5cm] (l11) {$U_1 \equiv L_1^{(1)}$}
            node[right of=l11, xshift=1cm] (l21) {$U_2 \equiv L_2^{(1)}$}
            node[below of=l11, yshift=0.15cm] (r1) {$R_1$}
            node[below of=l21, yshift=0.15 cm] (r2) {$R_2$}
            node[below of=r1] (l1) {$L_1$}
            node[below of=r2] (l2) {$L_2$}

            node[format, below of=l1, xshift=1.25cm, yshift=0.75cm] (a) {(a)}
            
            (l11) edge[blue] (l21)
            (l11) edge[blue] (r2)
            (l21) edge[blue] (r1)
            (r1) edge[gray] (l1)
            (r2) edge[gray] (l2)
            (l11) edge[gray, bend right] (l1)
            (l21) edge[gray, bend left] (l2)
            ;
        \end{scope}

        \begin{scope}[xshift=18cm, yshift=0cm]
            \path[->, thick]

            node[yshift=1.5cm] (l11) {$U_1 \equiv L_1^{(1)}$}
            node[right of=l11, xshift=1cm] (l21) {$U_2 \equiv L_2^{(1)}$}
            node[below of=l11, yshift=0.15cm] (r1) {$R_1$}
            node[below of=l21, yshift=0.15cm] (r2) {$R_2$}
            node[below of=r1, yshift=0cm] (l1) {$L_1^{(1)}$}
            node[below of=r2, yshift=0cm] (l2) {$L_2^{(1)}$}
            node[square, below of=r1, yshift=1cm] (fr1) {$r_1=1$}
            node[square, below of=r2, yshift=1cm] (fr2) {$r_2=1$}

            node[format, below of=l1, xshift=1.25cm, yshift=0.75cm] (a) {(b)}
            
            (l11) edge[blue] (l21)
            (fr1) edge[gray] (l1)
            (fr2) edge[gray] (l2)
            (l11) edge[blue] (r2)
            (l21) edge[blue] (r1)
            (l11) edge[gray, bend right=60] (l1)
            (l21) edge[gray, bend left=60] (l2)
            ;
        \end{scope}    

    \end{tikzpicture}
    }
	\end{center}
    \caption{(a) The block-parallel MNAR model, labeling missing variables as unmeasured for the purposes of constructing a SWIG; (b) The corresponding SWIG obtained from (a). }
    \label{fig:mnar-2}
\end{figure}
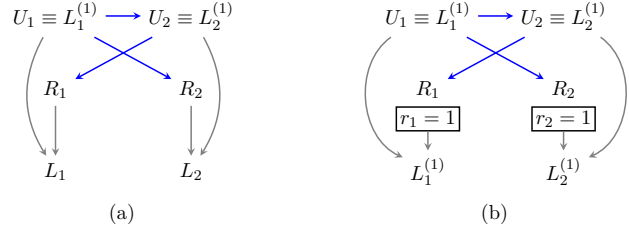

\subsubsection*{SWIGs do not make identification arguments for MNAR problems clearer}

The primary motivation for SWIGs in the causal inference context is to provide a graphical representation for independences that arise in identification arguments. The function of m-DAGs  is to provide precisely the same graphical representation in missing data problems, rendering SWIGs unnecessary. 

Take, for example, the independences that can be read from the m-DAG in Figure~\ref{fig:mnar-2}(a) and the corresponding SWIG in Figure~\ref{fig:mnar-2}(b). From either graph, we are able to extract independences of the form $R_k \ci L_k^{(1)}, R_{-k} \mid L_{-k}^{(1)}$, for $k \in \{ 1, 2 \}$ that define the binary block-parallel MNAR model. That is, the m-DAG itself is just as expressive as the SWIG for reading off independences important for identification in missing data models. In fact, we now argue that attempting identification using standard causal inference arguments from the SWIG may be problematic for missing data problems due to the single counterfactual nature of missing data, which as we discussed in Section~6 of our paper is more akin to the rank preservation assumption in causal inference.

To obtain the identification of the target law $p(l_1^{(1)}, l_2^{(1)})$ of the block-parallel MNAR model, we showed in Section~5.2 of our paper that a parallel use of the g-formula was required, which never arose in standard causal inference settings. If just the marginal $p(l_2^{(1)})$ is desired, this is obtained by marginalization. Now suppose instead, we attempted to perform identification arguments for  $p(l_2^{(1)})$ using the SWIG, in a manner similar to the derivation of the adjustment formula for causal inference. This would proceed as follows. First notice from the SWIG in Figure~\ref{fig:mnar-2}(b) that $L_2^{(1)} \ci R_2 \mid U_1, R_1$ -- this is analogous to conditional ignorability except $U_1$ is observed only when $R_1=1$. Thus we have,
\begin{align*}
    p(l_2^{(1)}) &= \sum_{r_1, u_1} p(r_1, u_1, l_2^{(1)}) \\
    & = \sum_{r_1, u_1} p(r_1, u_1) \, p(l_2^{(1)} \mid r_1, u_1) \\
    &= \sum_{r_1, u_1} p(r_1, u_1) \, p(l_2^{(1)} \mid r_1, u_1, r_2=1) \\
    &= \sum_{r_1, u_1} p(r_1, u_1) \, p(l_2 \mid r_1, u_1, r_2=1).
\end{align*}
Despite the final expression above appearing to be a function that is devoid of counterfactuals, the appearance of the variable $U_1$ in the expression  prevents identification, as $U_1$ is observed only when $R_1 = 1$. Since $R_1$ cannot be set to the value 1 in the formula above, we are unable to establish identification via the arguments presented. In fact, no sequential strategy that we are aware of based on the SWIG in Figure~\ref{fig:mnar-2}(b) would help establish identification. That is, the parallel fixing arguments presented in Section~5.2 of our paper are required here.

In short, while applying the SWIG construction to m-DAGs offers useful insights in causal problems with simple types of missingness, such as MCAR or MAR, we caution that such constructions must be done with care, as Levis and Kennedy have done.
 
\begin{figure}[t] 
	\begin{center}
	\scalebox{0.7}{
    \begin{tikzpicture}[>=stealth, node distance=2.25cm]
        \tikzstyle{format} = [thick, minimum size=5.0mm, inner sep=1pt]
        \tikzstyle{square} = [draw, thick, minimum size=4.5mm, inner sep=2pt]

        \begin{scope}[xshift=0cm, yshift=0cm]
            \path[->, thick]
            
            % node[] (x) {$X$}
            % node[right of=x] (a1) {$A^{(1)}$}
            node[] (a1) {$A^{(1)}$}
            node[above right of=a1, xshift=-0.25cm, yshift=-0.2cm] (x) {$X$}
            node[right of=a1, xshift=0.5cm] (y1) {$Y^{(1)}$}
            node[below of=a1, yshift=0.75cm] (ra) {$R_A$}
            node[below of=y1, yshift=0.75cm] (ry) {$R_Y$}
            node[below of=ra, yshift=1cm] (a) {$A$}
            node[below of=ry, yshift=1cm] (y) {$Y$}

            node[format, below of=a, xshift=1.25cm, yshift=1.2cm] (label) {(a)}

            (x) edge[blue] (a1)
            (a1) edge[blue] (y1)
            (x) edge[blue] (y1)
            (a1) edge[blue] (ry)
            (y1) edge[blue] (ra)
            (ra) edge[gray] (a)
            (ry) edge[gray] (y)
            (a1) edge[gray, bend right] (a)
            (y1) edge[gray, bend left] (y)
            ;
        \end{scope}

        \begin{scope}[xshift=6.5cm, yshift=0cm]
            \path[->, thick]
            
            % node[] (x) {$X$}
            % node[right of=x] (a1) {$A^{(1)}=A$}
            node[] (a1) {$A^{(1)}=A$}
            node[above right of=a1, xshift=-0.25cm, yshift=-0.2cm] (x) {$X$}
            node[right of=a1, xshift=0.5cm] (y1) {$Y^{(1)}=Y$}
            node[square, below of=a1, yshift=0.75cm] (ra) {$r_A=1$}
            node[square, below of=y1, yshift=0.75cm] (ry) {$r_Y=1$}
            node[below of=ra, yshift=1cm] (a) {\textcolor{gray}{$A$}}
            node[below of=ry, yshift=1cm] (y) {\textcolor{gray}{$Y$}}
    
            node[format, below of=a, xshift=1.25cm, yshift=1.2cm] (label) {(b)}

            (x) edge[blue] (a1)
            (a1) edge[blue] (y1)
            (x) edge[blue] (y1)
            (ra) edge[gray, dashed] (a)
            (ry) edge[gray, dashed] (y)
            (a1) edge[gray, bend right=60, dashed] (a)
            (y1) edge[gray, bend left=60, dashed] (y)
            ;
        \end{scope}

        \begin{scope}[xshift=14cm, yshift=0cm]
            \path[->, thick]
            
            % node[] (x) {$X$}
            % node[right of=x] (a1) {$A^{(1)}=A$}
            node[] (a1) {$A^{(1)}=A$}
            node[square, right of=a1, xshift=-0.7cm] (fa1) {$a$}
            node[above right of=a1, xshift=-0.cm, yshift=-0.2cm] (x) {$X$}
            node[right of=fa1, xshift=0.5cm] (y1) {$Y^{(1, a)}=Y^{(a)}$}
            node[square, below of=a1, yshift=0.75cm] (ra) {$r_A=1$}
            node[square, below of=y1, yshift=0.75cm] (ry) {$r_Y=1$}
            node[below of=ra, yshift=1cm] (a) {\textcolor{gray}{$A$}}
            node[below of=ry, yshift=1cm] (y) {\textcolor{gray}{$Y$}}

            node[format, below of=a, xshift=2cm, yshift=1.2cm] (label) {(c)}

            (x) edge[blue] (a1)
            (fa1) edge[blue] (y1)
            (x) edge[blue] (y1)
            (ra) edge[gray, dashed] (a)
            (ry) edge[gray, dashed] (y)
            (a1) edge[gray, bend right=60, dashed] (a)
            (y1) edge[gray, bend left=60, dashed] (y)
            ;
        \end{scope}

    \end{tikzpicture}
    }
	\end{center}
    \caption{(a) Extension of Levis and Kennedy's example to have two missing variables. (b) Graph obtained after fixing $R_A$ and $R_Y$ in parallel. (c) SWIG obtained by splitting the treatment variable $A$ after already having fixed $R_A$ and $R_Y$.}
    \label{fig:parallel}
\end{figure}
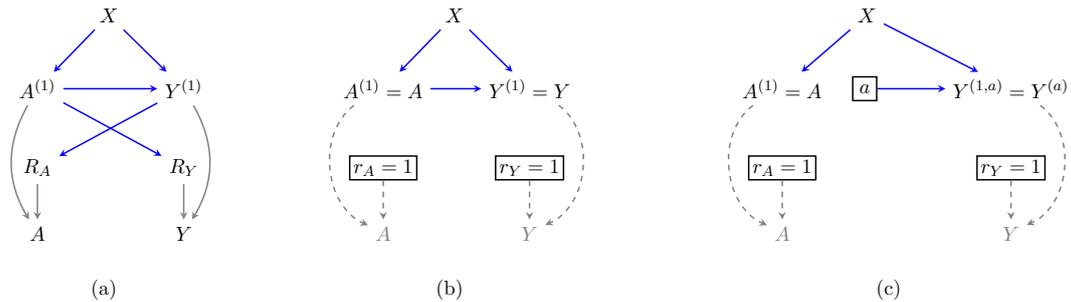

To illustrate, Figure~\ref{fig:parallel}(a) 
% presents an extension
extends the example of Levis and Kennedy
% 's example by introducing a causal inference problem featuring confounding, and 
to include missingness in both the outcome $Y$ and treatment $A$.  The goal here is to identify the average causal effect of $A$ on $Y$ in the absence of censoring.

Just as in the previous example shown in Figures~\ref{fig:mnar-2}(a, b), an argument based solely on SWIGs does not yield identification.  Instead, identification strategies that leverage restrictions encoded in both m-DAGs and SWIGs are necessary.  In particular, we first obtain the target law $p(X, A^{(1)}, Y^{(1)})$ by applying inverse probability weighting with the product of the propensity scores for $R_1$ and $R_2$, resulting in the graph in Figure~\ref{fig:parallel}(b). We then apply the g-formula on $A^{(1)}$ to adjust for confounding by $X$, taking advantage of the restrictions in the SWIG shown in Figure~\ref{fig:parallel}(c). This yields $p(Y^{(1,a)})$, the distribution of the outcome $Y$, had it not been censored and had treatment been set to value $a$, from which the average causal effect of interest may be obtained.

Finally, we note that there is currently no complete graphical identification theory for causal parameters associated with arbitrary m-DAGs that encode both MNAR and (possible) confounding by unmeasured common causes $U$.  We expect this theory to employ both SWIG and m-DAG constructions, as in the worked example of Levis and Kennedy, and our examples above.

%%%%%%%%%%%%%%%%%%%%%%%%%%%%%%%%%%%%%%%%%%%%%%%%%%%%%%%%%%%%
\subsection{On the discussion by Luo and Geng}

\cite{LuoGeng2025} extend our framework by focusing on self-censoring MNAR mechanisms, where the missingness of a variable depends directly on its unobserved value. They study such models with binary outcomes and establish identifiability results under explicit, testable conditions. Their discussion highlights how auxiliary variables, either baseline or follow-up measurements, can be leveraged to restore identification. In doing so, they illustrate that self-censoring structures, which were not emphasized in our paper, can still yield identification in important cases. They also point to future directions for extending these ideas beyond the binary case to more general settings. 

We thank Luo and Geng for their examples of self-censoring mechanisms. In our paper, we noted self-censoring as a form of MNAR but did not explore it further, since m-DAG models with self centering are never nonparametrically identified \citep{mohan13missing}. Since our emphasis was on nonparametric identification, we did not employ any restrictions not implied by the m-DAG factorization, including (i) relevant restrictions needed by instrumental variable or proxy methods, or (ii) constraints implied by state space restrictions. As \cite{LuoGeng2025}  illustrated, such assumptions can sometimes yield identifiability, and their results highlight concrete testable conditions under which this occurs.

We fully agree that self-censoring is of practical importance, with income surveys, sensitive health questions, and other social science settings providing common examples. Their discussion usefully demonstrates how auxiliary information, through baseline or follow-up variables, can be leveraged to address such scenarios.  

More broadly, we believe that in MNAR submodels that identify the target law, restrictions implied by the m-DAG always yield testable implications for the observed data law, with the exception of the permutation model \citep{robins97non-a}. Examples of such tests have been developed in prior work on graphical models \citep{mohan2014testability, nabi2022testability, guo2023sufficient, chen2023causal}. 

Overall, we view their discussion as complementary to our focus: while our framework aimed to catalog nonparametric identification results, their work illustrates how additional structure can both sharpen identifiability and connect to practical applications.

%%%%%%%%%%%%%%%%%%%%%%%%%%%%%%%%%%%%%%%%%%%%%%%%%%%%%%%%%%%%
\subsection{On the discussion by Wang and van der Laan}

% \cite{WangVdL2025} approach missingness through the lens of interventions, emphasizing the connection between missing data models and multivariate censoring. They argue that viewing missingness as a longitudinal intervention can yield efficient estimation strategies, particularly under sequential randomization assumptions, and they relate our framework to established censoring models in survival analysis. They also highlight how missing data DAGs can represent flexible MNAR submodels such as permutation missingness and block-conditional MAR, but caution that each specific m-DAG encodes strong local Markov restrictions that may be less plausible in practice. Their discussion raises questions about which MNAR submodels are most relevant in real-world applications and underscores the need to balance identifiability with plausibility and efficiency in estimation.

\cite{WangVdL2025}, like us, approach missingness through the lens of interventions. They emphasize the close relationship between missing data models and multivariate censoring models in survival analysis. They carefully prove that the permutation missingness model is MNAR rather than MAR and flesh out our example of why the permutation model is substantively plausible in some real world applications. They also discuss substantive settings where the block sequential model is plausible. 

They are less certain that the other identifiable MNAR models discussed in our paper are substantively plausible.  In our view, it is difficult to establish plausibility of \emph{any} missing data model in practice, whether it is formulated graphically or not.  One advantage of the sort of theory we present is a single formulation for a large class of identifiable models (including existing models in the literature such as those described in \citep{zhou10block,robins97non-a}).  Since no single model may be plausible in applications, an alternative strategy is to perform the analysis of interest under a wide class of identifiable missingness models, as a form of sensitivity analysis.
%, a point with which we largely agree.
% We return to the issue of the plausibility of graphical MNAR models below.

%\red{[FOR JAMIE!]}
% \section{On the %practical utility
% plausibility of identified missing data models}

We make one final observation regarding the plausibility of (or the lack there of) the identifiable  MNAR models considered in the paper.
 Identification in every case we discuss
 relies on an assumption that one or more counterfactuals $L_{j}^{(1)}$ suffice to control confounding of the ``effect'' of R$_{k}$ on $L_{k}$ where $j\neq k.$ But why should we privilege $L_{j}^{(1)}$ (or a set of such variables) as a set sufficient for adjustment?
%this be the case in practice? 
It seems much more plausible that there exist other unmeasured common causes of $R_k$ and $L_{k}$ (equivalently $L_{k}^{(1)}$) that would also need to be adjusted for to eliminate confounding. Are we assuming the $L_{j}^{(1)}$ suffice to control confounding because we believe it to be so or rather because we want to achieve identification for identification's sake?  
In fact, similar plausibility concerns arise in non-graphical identifiable models, such as non-monotone MAR.  In our view, plausible models of missingness feature a description of data generation that follows a temporal order (often representable as a DAG).  Unfortunately, to be realistic, even models of this type would feature unobserved confounding of the sort that would prevent identifiability.

%%%%%%%%%%%%%%%%%%%%%%%%%%%%%%%%%%%%%%%%%%%%%%%%%%%%%%%%%%%%
\subsection{On the discussion by Yang and Kim}

\cite{YangKim2025} emphasize the practical challenges of applying the m-DAG framework. They point out that while m-DAGs provide a principled way to encode assumptions and extend identification theory, their utility in practice depends critically on correct graph specification, which may be difficult to achieve in applied settings. They raise concerns about scalability in high-dimensional problems, where the number of variables and missingness patterns grows quickly, and about the feasibility of validating the conditional independence assumptions encoded in an m-DAG. They further stress the importance of developing systematic approaches to sensitivity analysis, noting that small misspecifications of the graph can have consequences for identification and downstream tasks. 

We thank Yang and Kim for raising important questions about the practical utility of m-DAGs, and graphical modeling more broadly. Our goal in this paper was primarily conceptual: to provide a general framework that connects causal and missing data perspectives and to catalog identification results that arise from this connection. Their emphasis on practical feasibility is appreciated, and complementary to our focus. That said, continuing work in the graphical modeling community over the last two decades has rendered much of the critiques of the graphical modeling approach to causal inference and missing data problems out of date.

Identification of causal and missing data parameters must, by necessity, rely on equality restrictions in the full data distribution.  Graphical models aim to encode these restrictions in a way that allows the construction of a plausible data generating mechanism consistent with a temporal order of events.  Models that obtain identifiability without a corresponding graphical representation, for instance the non-monotone missing at random (MAR) model, often lack such a plausible mechanism. Indeed, existing efforts that argue for the plausibility of models such as non-monotone MAR embed them into a graphical model or a mixture of such models \citep{robins97nonresponse}.

Over the last decade, an explosion of easy to use open-source packages that take advantage of graphical models have been developed for all tasks in causal inference and missing data, including establishing identification, constructing estimators and applying them to data, conducting sensitivity analyses, establishing bounds on non-identified parameters, and model selection.  A non-exhaustive list of these packages includes: 
Ananke \citep{lee2023ananke}, autobounds \citep{duarte2024automated},  dosearch \citep{tikka2021causal}, DAGitty \citep{textor16robust}, software developed as part of the Tetrad project (equipped with Python and R interfaces) \citep{ramsey2023py}, as well the pcalg package \citep{kalisch2012causal}, and extensions to deal with missing values \citep{andrews2024software}.

Further, sensitivity analyses results can be fruitfully applied to graphical models. For example, nongraphical results in \cite{robins99sensitivity} can easily be represented in graphical structures that encode MAR, conditional or sequential ignorability, and other types of Markov restrictions. For instance, the permutation model in \cite{robins97non-a} is shown to be a graphical model in our paper. 

In addition, graphical models allow a particularly powerful form of nonparametric sensitivity analysis, where consistent inferences are made in union models defined over a potentially large class of graphs that the analyst is uncertain about; see for example, \citep{vanderweele11new, yang2024statistical, shpitser10complete, shpitser10on, chang2024post, wang2025confidence}. 
For instance, the result in \citep{vanderweele11new} states that identification of the causal effect by covariate adjustment may be formulated without precise knowledge of the graph, but only via the set of common causes of the treatment and outcome, provided \emph{an} adjustment set exists.  The method developed by \cite{yang2024statistical} provide similar robustness guarantees with more complex types of identifying functionals, including the front-door functional, and the ratio functional arising in instrumental variable analysis.

We share the authors' view that algorithmic and computational advances will be needed to bring graphical identification methods to bear on high-dimensional problems with complex missingness patterns. We see this as an exciting frontier for future work, and one where continued integration of causal inference tools with missing data methodology will be especially fruitful. We believe smoothing and sparsity methods, as well as sum-product algorithm type methods developed in the graphical modeling literature may all be relevant for these advances \citep{lee2021markov}.

% \red{[FOR JAMIE!]}

%%%%%%%%%%%%%%%%%%%%%%%%%%%%%%%%%%%%%%%%%%%%%%%%%%%%%%%%%%%%
\section{Conclusion}
\label{sec:conc}

We thank the discussants for their thoughtful contributions. Together, their contributions point toward a roadmap: combine graphical identification, auxiliary information, intervention-based censoring perspectives, efficiency-oriented estimation, and structured sensitivity analysis into a practical toolkit for MNAR problems.

%%%%%%%%%%%%%%%%%%%%%%%%%%%%%%%%%%%%%%%%%%%%%%%%%%%%%%%%%%%%
%%%%%%%%%%%%%%%%%%%%%%%%%%%%%%%%%%%%%%%%%%%%%%%%%%%%%%%%%%%%
% \newpage
\bibliographystyle{plainnat}  % Change as needed (e.g., IEEE, APA)
\bibliography{references} 

%%%%%%%%%%%%%%%%%%%%%%%%%%%%%%%%%%%%%%%%%%%%%%%%%%%%%%%%%%%%
% \pagebreak 
% \appendix 

% \noindent {\LARGE \bf Appendix}
% \vspace{0.75cm}

% The appendix is organized as follows. \ldots 

% \vspace{0.75cm}
%%%%%%%%%%%%%%%%%%%%%%%%%%%%%%%%%%%%%%%%%%%%%%%%%%%%%%%%%%%%
% \section{Proofs}

\end{document}